\documentclass[journal]{IEEEtran}
\ifCLASSINFOpdf
\else
\fi

\usepackage{amssymb}

\usepackage{amsfonts}
\usepackage{latexsym}
\usepackage{amsmath}
\usepackage{array}
\newcolumntype{L}[1]{>{\raggedright\let\newline\\\arraybackslash\hspace{0pt}}m{#1}}
\newcolumntype{C}[1]{>{\centering\let\newline\\\arraybackslash\hspace{0pt}}m{#1}}
\newcolumntype{R}[1]{>{\raggedleft\let\newline\\\arraybackslash\hspace{0pt}}m{#1}}

\usepackage{hhline}
\usepackage[figuresright]{rotating}
\usepackage{lscape}
\usepackage{graphicx}
\usepackage{makecell}
\usepackage{multirow}
\usepackage{lipsum}
\usepackage{cuted}
\begin{document}
%
%
%
%
\makeatletter
\@addtoreset{footnote}{page}
\renewcommand{\thefootnote}{\ifcase\value{footnote}\or 1 \or 2 \or 3 \or * \fi}
\makeatother
\title{Energy-aware virtual machine selection method for cloud data center resource allocation}
\author{Nasrin~Akhter$^{1*}$,~\IEEEmembership{}
        Mohamed Othman$^{12*}$~\IEEEmembership{(Member,~IEEE),}
        Ranesh Kumar Naha$^3$~\IEEEmembership{}
}


\maketitle

\footnotetext[1]{Department of Communication Technology and Network, Universiti Putra Malaysia, 43400 UPM, Serdang, Selangor D.E., Malaysia.}
\footnotetext[2]{The author is an associate researcher at the Computational \mbox{Science} and Mathematical Physics Lab, Institute of Mathematical Science, Universiti Putra Malaysia, 43400 UPM, Serdang, Selangor D.E., Malaysia.}
\footnotetext[3]{School of Engineering and ICT, University of Tasmania, Hobart, TAS 7001, Australia.}
\footnotetext[4]{Corresponding author: nasrin786@gmail.com (Nasrin Akhter), \\ mothman@upm.edu.my (Mohamed Othman).}

%
%

\markboth{XXXXX,~Vol.~XX, No.~X, August~XXXX}%
{Akhter \MakeLowercase{\textit{et al.}}: }
%



\maketitle

\begin{abstract}
Saving energy is an important issue for cloud providers to reduce energy cost in a data center. With the increasing popularity of cloud computing, it is time to examine various energy reduction methods for which energy consumption could be reduced and lead us to green cloud computing. In this paper, our aim is to propose a virtual machine selection algorithm to improve the energy efficiency of a cloud data center. We are also presenting experimental results of the proposed algorithm in a cloud computing based simulation environment. The proposed algorithm dynamically took the virtual machines' allocation, deallocation, and reallocation action to the physical server. However, it depends on the load and heuristics based on the analysis placement of a virtual machine which is decided over time. From the results obtained from the simulation, we have found that our proposed virtual machine selection algorithm reduces the total energy consumption by 19\% compared to the existing one. Therefore, the energy consumption cost of a cloud data center reduces and also lowers the carbon footprint. Simulation based experimental results show that the proposed heuristics which are based on resource provisioning algorithms reduce the energy consumption of the cloud data center and decrease the virtual machine's migration rate.
\end{abstract}


%
\IEEEpeerreviewmaketitle

\section{Introduction}
%
%
%
%
\IEEEPARstart{T}{he} power consumption of distributed and large scale systems, like the grid and cloud data center, has enormously increased its operational cost and its impact on the environment. Indeed, they need a massive electrical power supply which is a major concern for various institutions. The data center’s maximum power spent on underutilized servers and on the cooling systems are used to cool off the underutilized servers. To facilitate cloud services, we need to build a large scale data center with over a thousand physical nodes, which requires a large amount of electrical power. With the growing demand of cloud computing, energy consumption will be vastly increased in the near future. The power efficiency of hardware and proper resource management helps to minimize electrical energy costs. A recent study shows that servers usually operate from 10\ to 50\% of their full capacity, which has been concluded from the collected data from 5000 operational servers over a half year period [1]. At the same time, a completely idle server consumes its maximum power at its peak.

A Cloud computing environment is built using one or more data centers, and these data centers have many computing resources which we call servers or hosts. Multiple virtual machines could be allocated for every host through virtual technology, and each virtual machine acts as an individual physical machine with their own OS and system resources. As a concept, cloud computing users are able to use computing resources as payment, because customer dissatisfaction should be avoided through SLA. When CPU utilization exceeds its limit due to an oversubscription agreement,  SLA terms could be violated. However, a VM migration before a possible oversubscription is a viable solution to avoid SLA violation [35]. On the other hand, energy saving is possible through VM consolidation by defining under-loaded hosts. In both situations, it is necessary to choose VMs for migration and also a new placement is needed for the VMs to be migrated.

In this paper, we have described VM placement and selection models. We have proposed a new VM selection method. The proposed method works by analysing RAM and network bandwidth usages. It reduces energy consumption in cloud data centers and we developed a new VM selection algorithm. A Simulation was performed under a complex scenario, and the results from the simulation are presented in this paper. Finally, we have demonstrated an analysis on obtained simulation results. The final results have shown that our proposed algorithm is more energy efficient compared to the existing one.


\section{Related Work}
\label{Rela_wo}
The promising model for utility computing was proposed by [2,3] which delivers various cloud services. Utility computing provides computation, storage, data access, software, and other software services. The increasing demand of cloud computing power consumption for data centers and cooling system have raised tremendously. However, reducing power consumption of cloud computing infrastructures is a challenging research issue.\\

Power management of virtualized data center in terms of energy-aware had been studied by Nathuji and Schwan [4]. Cloud computing resource provision is unpredictable and workload has varied over time. Random online algorithms and non-deterministic online algorithms normally improve the quality in a similar scenario compared to deterministic algorithms, as discussed by Ben-David et al. [5]. Algorithm depends on the input provided from distributed model and cannot be modeled using a plain statistical distribution [6,7] due to the complexity of realistic world setting. Determining the resource usage through application is not easy modelled through plain probability distribution which is shown by several studies [6,8,9].\\

A new operating system, ``Muse'' was proposed by Chase et al. [10] which can be used in a hosting center, and it is an adaptive resource management system which plays a vital role in integrating energy and power resources for the hosting center. The Muse approach considers resources such as CPU servers by managing other resources like network bandwidth, disks, and memory energy saving, which is also possible [11,12,13]. Pinheiro et al. [14] introduced three “double-threshold VM selection policies” to determine whether VMs can be migrated. The basis of this policy is to set a lower and upper utilization threshold, so all VMs will be allocated to the host depending on these thresholds.\\

Beloglazov and Buyya [15] had proposed an efficient resource management policy through VM consolidation for a cloud based data center, and they showed an overall operational cost reduction by using their proposed algorithm. Afterward, Beloglazov et al. [16] had proposed a high-level system architecture for cloud data centers by introducing the green service allocator and  proposed an VM allocation algorithm by using a modified Best Fit Decreasing (BFD) [17] algorithm. This modified algorithm works based on the current CPU utilization. Beloglazov and Buyya [18] proposed several algorithms for VM allocation; these algorithms find overloaded and under loaded hosts, and VM from overloaded hosts which are migrated to a new location where the host is under loaded. They had proposed three policies, Minimum Migration Time (MMT), Random Choice (RC) and Maximum Correlation (MC) for VM selections, basically these algorithms select overloaded hosts repeatedly until the host is considered to be overloaded. The MMT, RC and MC policies were developed based on the idea which had been proposed by Verma et al. [19]. \\

Cao et al. [20] had proposed a power-saving approach based on a demand forecast for the allocation of VMs. They will try to reduce the total CPU frequency by switching the hosts on/off. The authors took two types of resources into consideration: computing cores and memory. Naha et al. [21, 22, 34] proposed the cloud brokering and load balancing method for the cloud data center, the work was validated further in [33], but they did not consider the energy-awareness issue. We had proposed Energy aware VM selection Algorithm for the cloud data center in [23]. Raycroft et al. [24] analyzed the energy consumption of the global VM allocation using various real world policies under a realistic testing scenario. However, the simulation was centered on a single application, and it does not take into consideration the communication among VMs across regions. \\

Wu et al. [25] proposed a scheduling algorithm by using a dynamic voltage frequency scaling (DVFS) technique for the cloud data center. However,  during VM consolidation and VM migration DVFS did not improve the power consumption as Beloglazov and Buyya had shown. [18]. An energy efficient scheduling of virtual machines (EEVS) algorithms was proposed by Ding et al. [26] which had reduced the energy consumption of cloud data center, however, they overlooked the VM migration and transitions of the processor. Wolke et al. [27] found that periodic reallocations and combinations of the controller’s placement achieved the highest energy efficiency with predefined service levels. A detail review on energy aware resource allocation of cloud data center were presented in our previous work [32].

\section{VM migration and consolidation}
\label{VM_mig_con}
Power consumption of the cloud data center had increased because of an underutilized host and inefficiency of the resource management. On the other hand, overutilization of a server is caused by a system failure and this can increase SLA violation. To reduce overutilization, we may need to migrate VMs from an overutilized host to an underutilized host. Managing resources efficiently is an interesting issue but VM consolidation is one of the most important solutions for this problem. The energy consumption of a data center can be reduced by a live VM migration and consolidation.

\subsection{The single VM migration problem}
On a single host or a physical server, multiple VMs can be allocated. In terms of energy and performance awareness, a dynamic VMs consolidation problem and time discretion can be divided into several time frames, where each frame takes one second. The resource service provider pays the energy cost which is consumed by the physical servers. The cost will be calculated by multiplying the cost per unit energy and the time period. The resource usage and resource capacity of a single host is characterised by the CPU usages.\\

Though VMs can experience a dynamic workload during an operation, CPU usages varies over time. When the maximum CPU performance that is allowed exceeds, the host is considered to be oversubscribed. In this case, the established SLA policy between the service provider and consumer will be violated. The service provider will need to pay a penalty for violating the SLA and the penalty is calculated by multiplying the violated time frame of the SLA violation with the cost per unit. To solve the oversubscription problem, a single VM should be migrated to another host. This migration process will decrease the CPU’s utilization and will help to maintain the utilization threshold. However, during the migration, if another host that is used for migration in that specific time, the energy cost will be doubled. Defining when a migration should be initiated is quite challenging, especially when the minimization of energy cost and SLA violation costs take place. \\

\begin{figure*}
	\centering
	\includegraphics[width=4.5in]{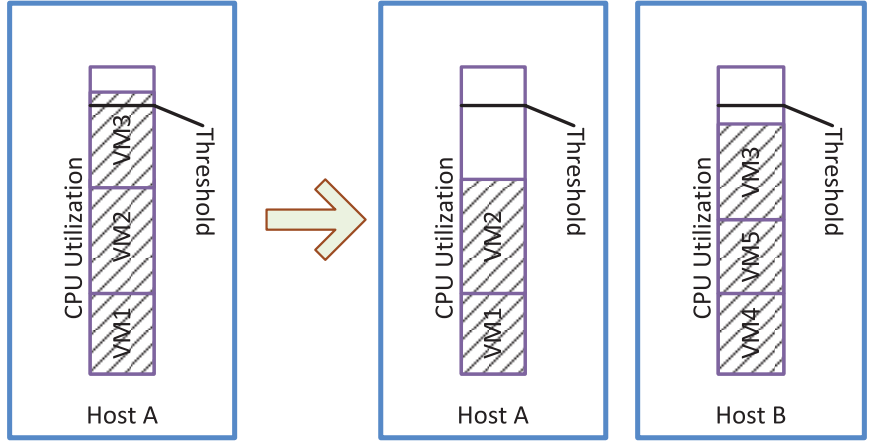}
	\caption{Scenario of single VM migration.}
	\label{fig_ssVMm}
\end{figure*}

Fig. 1 shows how a single VM migration helps to protect a possible system failure before the host is heavily overloaded using  the CPU utilization threshold. The analysis of cost function, optimal offline and online algorithm has been performed by Beloglazov and Buyya [18]. The cost of consumed energy by the server is paid by the resource provider and it is calculated as \(C_pt_p\) (\(C_p\) is the power cost and \(t_p\) is time period). The resource capacity of a host is used by VMs, which is  characterised by the CPU’s performance parameter. Due to dynamic workload, the CPU usage varies over time and it its usage is considered to be oversubscribed when the request of VMs exceeds the CPU's maximum performance. The oversubscription issue occurs when the SLA is violated, and the penalty for SLA violation is calculated by \(C_vt_v\) (\(C_v\) is the SLA violation cost of single unit of time and \(t_v\) is the SLA violation time duration). During the VM migration an extra host is accommodated until the migration process is completed and the time needed for migration is \(T\) so, total power consumption during migration is \(2C_pT\). At some point of time \(v\), if SLA violation starts and SLA violation continued until \(m\), then the total SLA violation time \(r\) will be as Eq. (1). \\

\begin{equation}\label{first}
r=m-v
\end{equation}

The cost function \(C(v,m)\) for SLA violation will be as Eq. (2) for three different cases.

\begin{strip}
\begin{equation}\label{2nd}
\footnotesize
C(v,m)=\left\{\begin{array}{l@{\quad}l} (v-m)C_p & \text{if} \ m<v,v-m \geq T, \\ (v-m)C_p+2(m-v+T)C_p+(m-v+T)C_v & \text{if} \ m \leq v,v-m<T, \\ rC_p+(r-m+v)C_p+rC_v & \text{if} \ m>v. \end{array} \right.
\end{equation}
\end{strip}

In the first case, VM migration starts before SLA violation \((m < v)\) and the migration starts when the SLA is violated \((v - m \geq T)\). In this case, the  duration of SLA violation is 0. In the second case, VM migration starts before the SLA violation , after which the migration starts later, even though the SLA has been violated. Lastly, in the third case, VM migration starts after the SLA violation. According to Beloglazov and Buyya, [18], ``The optimal offline algorithm for the single VM migration problem incurs the cost of \(\frac{T}{s}\), and is achieved when \(\frac{(v-m)}{T}=1\) and the competitive ratio of the optimal online deterministic algorithm for the single VM migration problem is \(2 + s\), and the algorithm is achieved when \(m = v\)''.

\subsection{Problem of Dynamic VM Consolidation}
The resource utilization of cloud data centers could be improved by dynamic VM consolidation, which can also improve energy efficiency. Dynamic VM consolidation can determine when VMs reallocation should be initiated for an overloaded host. The VMs reallocation decision-making influences the proper resource utilization and QoS requirements delivered by the system.\\

In our work, we deal with complex problems of dynamic VM consolidation, which requires us to consider multiple VMs and multiple hosts. All VMs confront variable workloads. Every host has a maximum CPU capacity limit and VM could be allocated within this limit. Based on heuristics, if the CPU capacity limit is exceeded, then VMs can be migrated through live migration. It is assumed that when a host is in idle mode, it considered to be in switch off mode and it consumes no power. All functioning hosts are referred to be in active mode. However,  the problem is when a VMs should be migrated for minimizing power consumption. We will propose certain algorithms which are chosen for the appropriate VMs migration to minimize power consumption.\\

\begin{figure*}
	\centering
	\includegraphics[width=3in]{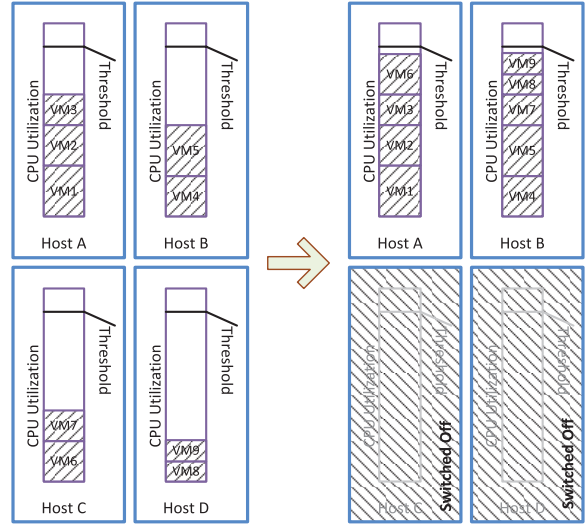}
	\caption{Idle hosts turned to switched off after dynamic VM consolidation.}
	\label{fig_ssVMm}
\end{figure*}

Fig. 2 shows a dynamic VM consolidation which saves power consumption by using VM consolidation and by turning off the idle host after VM consolidation. For a dynamic VM consolidation, we have assumed that there are n homogeneous hosts with an \(Ah\) capacity for each host and the maximum capacity of the CPU that can be allocated is \(Av\). Hence, the maximum capacity of a host for VM allocation is \(m=\frac{A_h}{A_v}\). Thereofore, the total number of VMs that will be allocated to all hosts is \(mn\). We assume that an idle host consumed no power because no VM is allocated in  an idle host, so it is switched off or in sleep mode working under a negligible power consumption. The total cost \(C\) for active host is shown in Eq. (3).

\begin{equation}\label{first}
C=\sum_{t=t_0}^T \left( C_p\sum_{i=0}^n a_{ti} + C_v \sum_{j=0}^n v_{tj}\right)
\end{equation}

In Eq. (3), initial time is \(t_0\); the total time is \(T\), \(a_{ti}\) indicating whether the host \(i\) is active at the time \(t\); \(v_{tj}\) indicating whether the host \(j\) is experiencing an SLA violation at the time \(t\).
The dynamic VM consolidation upper bound of the competitive ratio of the optimal online deterministic algorithm (ALG) comparing with optimal offline algorithm (OPT) is shown in Eq. (4).

\begin{equation}\label{first}
\frac{ALG(I)}{OPT(I)} \leq 1+\frac{ms}{2(m+1)}
\end{equation}

\subsection{Heuristics for VM Consolidation}
The CPU utilization threshold is calculated by analysing the historical data which is based on the resource usages by the VMs. The heuristics based analysis improves the decision making for the service allocation. The Heuristics algorithm automatically adjusts the utilization threshold with the help of a statistical analysis on the historical data which gathered during VMs lifetime. A previous study shows that the heuristics algorithm improves the energy consumption and service the quality [18]. The VM consolidation using historical CPU utilization had continued throughout the simulation as illustrated in Fig. 3.

\begin{figure*}
	\centering
	\includegraphics[width=3in]{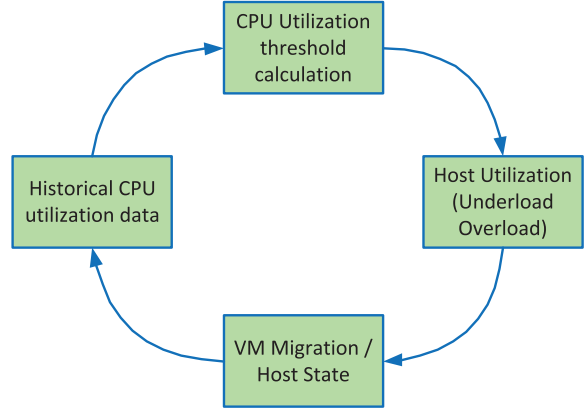}
	\caption{Process of dynamic VM consolidation.}
	\label{fig_ssVMm}
\end{figure*}

\subsection{VMs  Placement and Selection in cloud}
Several VMs could be ran in a single machine based on the service request. Under the virtualization concept, multiple operating systems could be ran on multiple VMs in a single physical machine. Through the VM consolidation we can use system resources efficiently and unused resources could be placed in lower power state to save energy. Several heuristic methods have been proposed for VM consolidation by Beloglazov and Buyya [18]. To perform a VM consolidation it is necessary to select an overloaded host first and then, we must follow specific criteria for host overloading. Furthermore, one or more hosts need to be migrated into an underloaded host or a new host. Depending on the load, several underloaded hosts could be migrated in one or more hosts. In this case, surely some underloaded host could be programmed in sleep mode after migrating VMs, which is running on them. Which VM and what type of VM should be migrated first is the another an issue which should be taken into consideration. Furthermore, the selected VM will be placed into a new location.

\section{Algorithm for resource optimization}
In an operational environment, a thorough knowledge on future events is not imaginable through the control of algorithms. Such events deal with online problems. The optimization issue where input is received in an online manner and output is also produced online and is called an online problem, according to [7]. The algorithm developed for these types of problems is called an online algorithm. To characterise the efficiency and performance of these algorithms, we can apply a competitive analysis. Based on the online algorithm knowledge, a competitive analysis generates the worst possible input. The input of online algorithm maximizes the modest ratio which is based on the outside world. This input should not be confused with the internal states of algorithms such as internal memory and control.

\section{Proposed method for VM Selection}
In this section we are going to propose a new policy for VM selection. Our proposed policy finds the appropriate VMs for migration.

\begin{table}[!t]
	\renewcommand{\arraystretch}{1.3}
	\caption{Energy Aware VM Selection (EAVMS).}
	\label{table_example}
	\begin{tabular}{l}
		\hline
		\textbf{Algorithm 1:}\\
		\hhline{=}
		1 \hspace{.20cm} Input: vmList from host \hspace{.3cm} Output: Selected VM\\
		2 \hspace{.20cm} mVms $\leftarrow$ get MVms(host) \\
		3 \hspace{.20cm} if mVms is NULL then \\
		4	\hspace{.26cm} $\vert$	\hspace{.3cm}	return null \\
		5 \hspace{.20cm} else \\
		6	\hspace{.26cm} $\vert$	\hspace{.3cm}	vmToM $\leftarrow$ null \\
		7	\hspace{.26cm} $\vert$	\hspace{.3cm}	nM $\leftarrow$ Double.MAX\_VALUE \\
		8	\hspace{.26cm} $\vert$	\hspace{.3cm}	foreach vm in mVMs do \\
		9	\hspace{.26cm} $\vert$	\hspace{.3cm}	$\vert$	\hspace{.3cm} if vm.isInM() $\leftarrow$ False then \\
		10	\hspace{.05cm} $\vert$	\hspace{.3cm}	$\vert$	\hspace{.3cm} $\vert$	\hspace{.3cm} metric $\leftarrow$ vm.getRam() \\
		11	\hspace{.05cm} $\vert$	\hspace{.3cm}	$\vert$	\hspace{.3cm} $\vert$	\hspace{.3cm} if m < mM then \\
		12	\hspace{.05cm} $\vert$	\hspace{.3cm}	$\vert$	\hspace{.3cm} $\vert$	\hspace{.3cm} $\vert$	\hspace{.3cm}  mM $\leftarrow$ m \\
		13	\hspace{.05cm} $\vert$	\hspace{.3cm}	$\vert$	\hspace{.3cm} $\vert$	\hspace{.3cm} $\vert$	\hspace{.3cm}  vmToM $\leftarrow$ vm \\
		14 \hspace{.05cm} $\vert$	\hspace{.3cm} return vmToM\\
		\hline
	\end{tabular}
\end{table}

\subsection{General Description}
The dynamic VM consolidation migrates VMs when a host is considered to be underloaded or overloaded. An underloaded host is placed into sleep mode after migrating all running VMs on it. In the case of an overloaded host, at first, it is necessary to select VMs that need to be migrated. After that, the system needs to find a new placement for migrating all the selected VMs. Our new VM Selection policy decides which VMs needs to migrate first. This algorithm migrates VMs until a a host is considered to be overloaded.

\subsection{Problem Formulation}
Our Maximum Migration Time (MxMT) policy migrates VM which took the maximum time for completing the migration. In a host, several VM could be allocated. When a host is considered to be overloaded, the MxMT policy will choose VM which need the longest period for migrating compared to other VMs that are allocated for that host. The selected VM that is chosen for the migration denoted as  \(v\). The migration time is defined by dividing the available bandwidth for the host \(h\) from the amount of used RAM by the VM . The proposed MxMT policy finds the VM which satisfies the following condition:

\begin{equation}\label{First}
v \in V_h | \forall_x \in V_h, \frac{RAM_u(v)}{NET_h} \geq \frac{RAM_u(x)}{NET_h},
\end{equation}

The proposed algorithm checks the condition stated in Eq. (5), for all VMs our method finds the VM which took maximum time frame for migration. In Equation (5.1), \(RAM_u(v)\) we find the amount of utilized RAM used by VM \(v\). Similarly, \(RAM_u(x)\) is the utilized RAM by VM \(x\). \(NET_h\) is the unused network bandwidth which is unoccupied for the host \(h\).

\begin{table}[htbp]
	\centering
	\caption{Power consumption of the servers at 0\% to 50\% load level (kWh).}
	\label{PowerModel}
	\begin{tabular}{C{2cm}|L{0.5cm}L{0.5cm}L{0.5cm}L{0.5cm}L{0.5cm}L{0.5cm}} \hline
		\textbf{Server}&\textbf{0\%}&\textbf{10\%}&\textbf{20\%}&\textbf{30\%}&\textbf{40\%}&\textbf{50\%} \\ \hline
		HP ProLiant G4 & 86 & 89.4 & 92.6 & 96 & 99.5 & 102 \\ \hline
		HP ProLiant G5 & 93.7 & 97 & 101 & 105 & 110 & 116 \\ \hline
		\hline\end{tabular}
\end{table}

\begin{table}[htbp]
	\centering
	\caption{Power consumption of the servers at 60\% to 100\% load level (kWh).}
	\label{PowerModel}
	\begin{tabular}{C{2cm}|L{0.5cm}L{0.5cm}L{0.5cm}L{0.5cm}L{0.5cm}} \hline
		\textbf{Server}&\textbf{60\%}&\textbf{70\%}&\textbf{80\%}&\textbf{90\%}&\textbf{100\%} \\ \hline
		HP ProLiant G4 & 106 & 108 & 112 & 114 & 117 \\ \hline
		HP ProLiant G5 & 121 & 125 & 129 & 133 & 135 \\ \hline
		\hline\end{tabular}
\end{table}

\section{Algorithm for Virtual Machine Selection}
The pseudocode of our proposed Energy Aware VM Selection (EAVMS) algorithm is presented in Table 1. The proposed algorithm select the VM for migration from overloaded host. The key policy of this algorithm is that it selects the VM that requires the maximum time frame to migrate compared to the migration time of other allocated VMs to a specific host. The migration time is estimated by dividing the available network bandwidth to the amount of current RAM used by the VM.

\section{Experimental and Simulation Setup}
For our simulation we have used CloudSim 3.0 toolkit [28] which is a modern simulation tool that supports the modelling of cloud data center with on demand virtualization resources and application management. We modeled a data center with eight hundred heterogeneous physical machines and over 1000 running VMs in a simulation environment.

\subsection{System Power Utilization Model}
The hosts' power consumption is defined according to the power consumption of HP ProLiant G4 and G5 server. According to the power consumption of these servers, a server consumes from 86 W with 0\% CPU utilization and maximum 135 W with 100\% CPU utilization. Power consumption on different levels of utilization for these two servers is shown in Table 2 and 3. The dual core server was chosen for this simulation because it is easy to overload servers using lesser workload. On the other hand, dual cores CPUs are adequate for the evaluation of resource management related to algorithms which are designed for multicore CPU architecture.

\subsection{Network Behaviour Modeling}
CloudSim [28] simulation framework supports the modeling of realistic networking topologies and models. For internetworking cloud entities such as hosts, data centers and connected end users are based on the conceptual networking abstraction model. This model is based on a latency matrix instead of actual network devices like routers and switches. CloudSim is an event based simulation framework, event management engine of CloudSim which maintains a latency for transmitting messages within the cloud entities. The number of network nodes are stored in a topology description table in BRITE [31] format. These nodes represent all CloudSim entities including hosts, data center, and cloud brokers. BRITE information is loaded every time during CloudSim initialization and is used for generating latency matrix.



\subsection{Workload Characterization}
Workload traces form the Real system are acceptable for evaluation a simulation. Workload data has been taken from PlanetLab [29] which is a part of CoMon [30] project. During a random period of ten days, the Workload traces have collected from March 2011 to April 2011. The CPU utilization is below 50\% in terms of workload traces, and during the simulation the VM assignment had been random. Table 4 represents the characteristics of the workload. A CPU utilization had been collected from the servers that were located at more than 500 places around the world. Thousands of VMs were deployed in the workload as shown in Table 4 and measurements of CPU utilization was 5 minutes.

\begin{table*}[htbp]
	\centering
	\caption{CPU utilization of workload data [18].}
	\label{workload}
	\begin{tabular}{L{2.2cm}|C{1.7cm}L{1.4cm}L{1.4cm}C{1.7cm}C{1.5cm}C{1.7cm}} \hline
		\textbf{Date}&\textbf{Number of VMs}&\textbf{Mean}&\textbf{St. dev.}&\textbf{Quartile 1}&\textbf{Median}&\textbf{Quartile 3} \\ \hline
		03/03/2011 & 1052 & 12.31\% & 17.09\% & 2\% & 6\% & 15\% \\ \hline
		06/03/2011 & 898 & 11.44\% & 16.83\% & 2\% & 5\% & 13\% \\ \hline
		09/03/2011 & 1061 & 10.70\% & 15.57\% & 2\% & 4\% & 13\% \\ \hline
		22/03/2011 & 1516 & 9.26\% & 12.78\% & 2\% & 5\% & 12\% \\ \hline
		25/03/2011 & 1078 & 10.56\% & 14.14\% & 2\% & 6\% & 14\% \\ \hline
		03/04/2011 & 1463 & 12.39\% & 16.55\% & 2\% & 6\% & 17\% \\ \hline
		09/04/2011 & 1358 & 11.12\% & 15.09\% & 2\% & 6\% & 15\% \\ \hline
		11/04/2011 & 1233 & 11.56\% & 15.07\% & 2\% & 6\% & 16\% \\ \hline
		12/04/2011 & 1054 & 11.54\% & 15.15\% & 2\% & 6\% & 16\% \\ \hline
		20/04/2011 & 1033 & 10.43\% & 15.21\% & 2\% & 4\% & 12\% \\ \hline
		\hline\end{tabular}
\end{table*}

\subsection{Experimental Setup}
Each node of data center is modelled with a dual core CPU and the performance of each core is equivalent to 1860 MIPS for HP ProLiant ML110 G4 server and 2660 MIPS for HP ProLiant ML110 G5 servers. Each server had been modelled with 1 GbPS network bandwidth. The types of VM were exhibited as Amazon EC2 instances and were listed in Table 5. The VMs of the data center were deployed with single core CPU because the workload data comes from single core VMs. Initially, system resources were occupied as VM types by the VMs. However, during the life time simulation, fewer system resources that were occupied had been consistent with workload data for a dynamic VM consolidation.

\begin{table}[htbp]
	\centering
	\caption{Configuration of instances.}
	\label{Instances}
	\begin{tabular}{L{3cm}|C{2cm}|C{2cm}} \hline
		\textbf{Instance Type} & \textbf{CPU Speed MIPS} & \textbf{Storage GB} \\ \hline
		High-CPU Medium Instance & 2500 & 0.85  \\ \hline
		Extra Large Instance & 2000 & 3.75  \\ \hline
		Small Instance & 1000 & 1.7  \\ \hline
		Micro Instance & 500 & 613  \\ \hline
		\hline\end{tabular}
\end{table}

\subsection{Simulation Scenario}

The modelled data center is connected to the internet and user requests are generated from the internet. User requests are generated according to the workload traces and are passed to the data center. The data center processes requests, consolidates and deconsolidates VMs when necessary. Figure 4 shows the basic scenario of simulation for our proposed algorithms.

\begin{figure*}
	\begin{center}
		\includegraphics[width=0.6\textwidth]{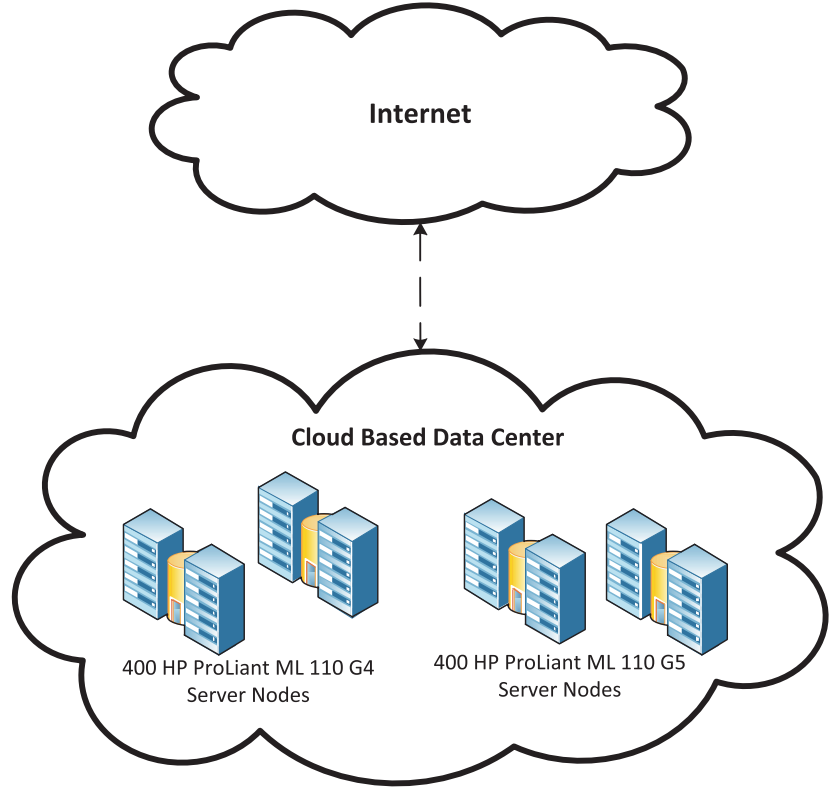}
		\caption{Scenario of the Simulation.}
		\label{fig:3 Scenario}       
	\end{center}
\end{figure*}

\subsection{Physical Resources}
We ran a simulation with an arrangement on three of the same type physical machines. The configuration of the physical machines are Intel$^{\circledR}$  \ core\textsuperscript{TM} 2 Duo CPU E8400 3.00 GHz processor and with 500GB storage, 4 GB Ram, Windows 7 with 32-bit OS and with 250 GB storage.

\section{Performance Metrics}
The Performance Measured had been established by following the three main performance metrics. These metrics are proposed by Beloglazov and Buyya [18].

\subsection{Energy Consumption}
The total energy consumption is measured by taking into consideration the total energy consumption made by a data center during the application workload. The unit for energy consumption is kilo watt per hour (kWh).

\subsection{SLA Violation}
The SLA violation (SLAV) percentage is defined as the percentage of SLA violation events which are relatively in accordance to the total number of the processed time frames. SLA violation is calculated through Performance Degradation due to Migration (PDM) and SLA Time per Active Host (SLATAH).

\subsection{Number of VM Migrations}
The number of VM migrations is initiated by the VM manager during the adaptation of the VM placement. We calculate number of VMs migrated during the simulation with single day workload. Each time during our simulation, we simulate with a 24-hours workload.

\subsection{Energy and SLA Violations}
The objective of proposing new algorithms is to reducie the violation of SLA and energy consumption. Therefore, Energy and SLA Violation (ESV) in Eq. (6) are used as a combined metric.

\begin{equation}
ESV = E.SLAV
\end{equation}

\section{Simulation Results and Discussions}

We have simulated our proposed algorithm in a simulated environment. The simulation was conducted using CloudSim toolkit with real life workload which had been derived from thousands of PlanetLab physical servers which were located around the world. During the simulation, we had combined our proposed algorithm with the previously proposed best algorithm combinations proposed by [18]. The best algorithm combinations are THR-MMT-1.0, THR-MMT-0.8, IQR-MMT-1.5, MAD-MMT-2.5, LRR-MMT-1.2 and LR-MMT-1.2.

\clearpage

\begin{figure*}
	\begin{center}
		\includegraphics[width=0.8\textwidth]{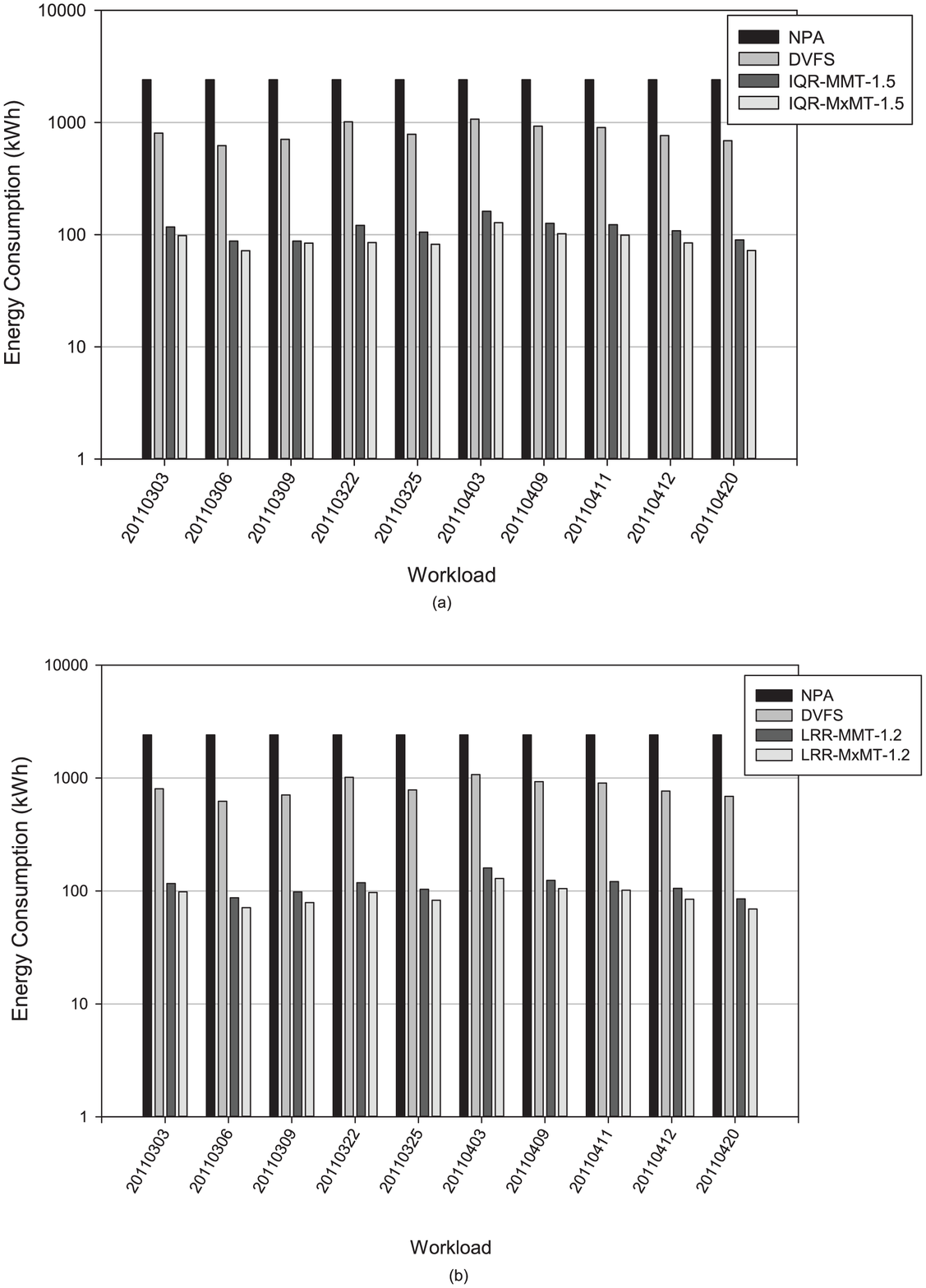}
		\caption{Result Comparison of Energy Consumption (a) Maximum Migration Time with IQR (b) Maximum Migration Time with LRR.}
		\label{F2EnergyIQRLRR}       
	\end{center}
\end{figure*}

\begin{figure*}
	\begin{center}
		\includegraphics[width=0.8\textwidth]{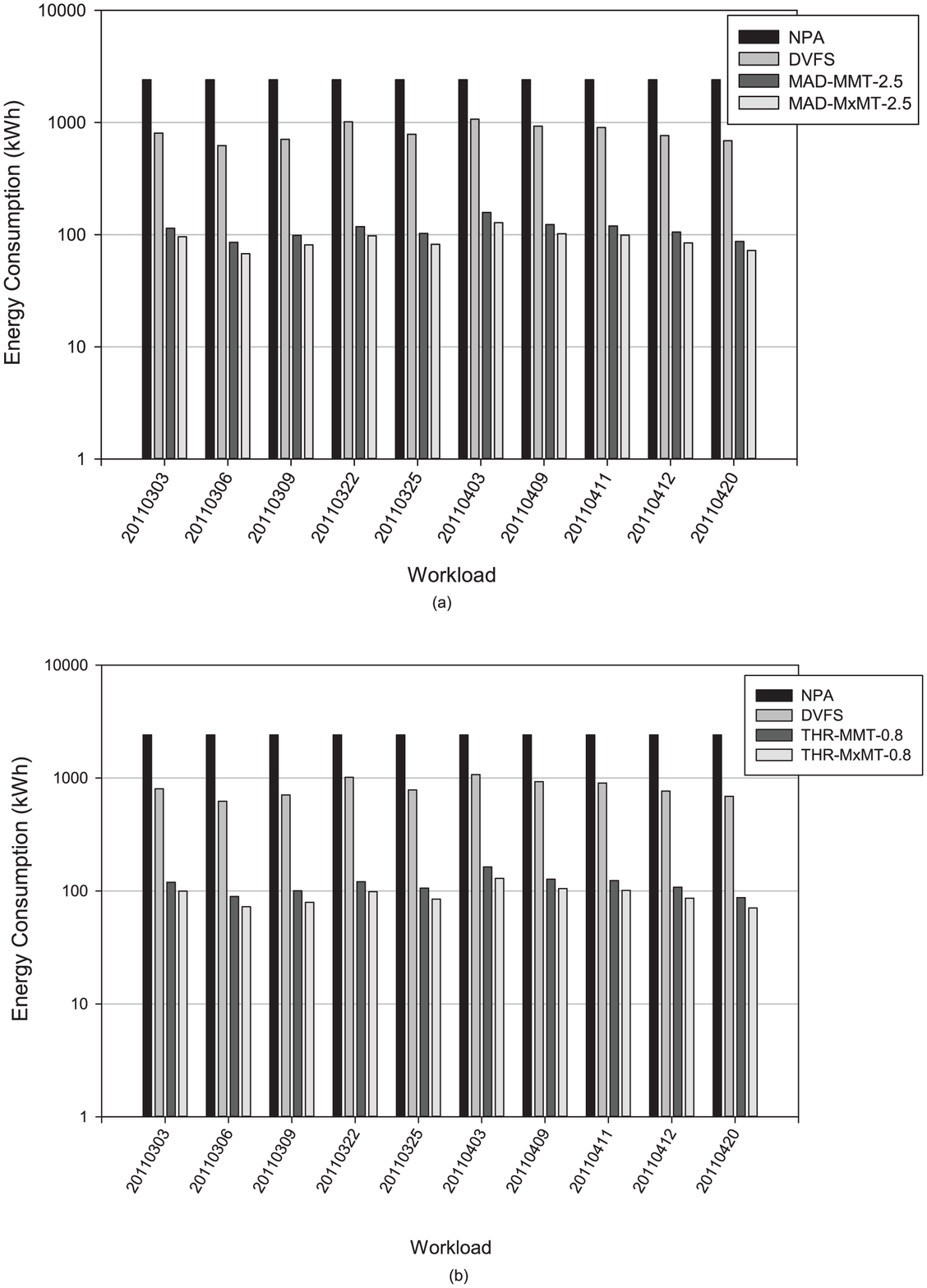}
		\caption{Result Comparison of Energy Consumption (a) Maximum Migration Time with MAD (b) Maximum Migration Time with THR 0.8.}
		\label{F2EnergyMADTHR}       
	\end{center}
\end{figure*}

\clearpage

\begin{figure*}
	\begin{center}
		\includegraphics[width=0.8\textwidth]{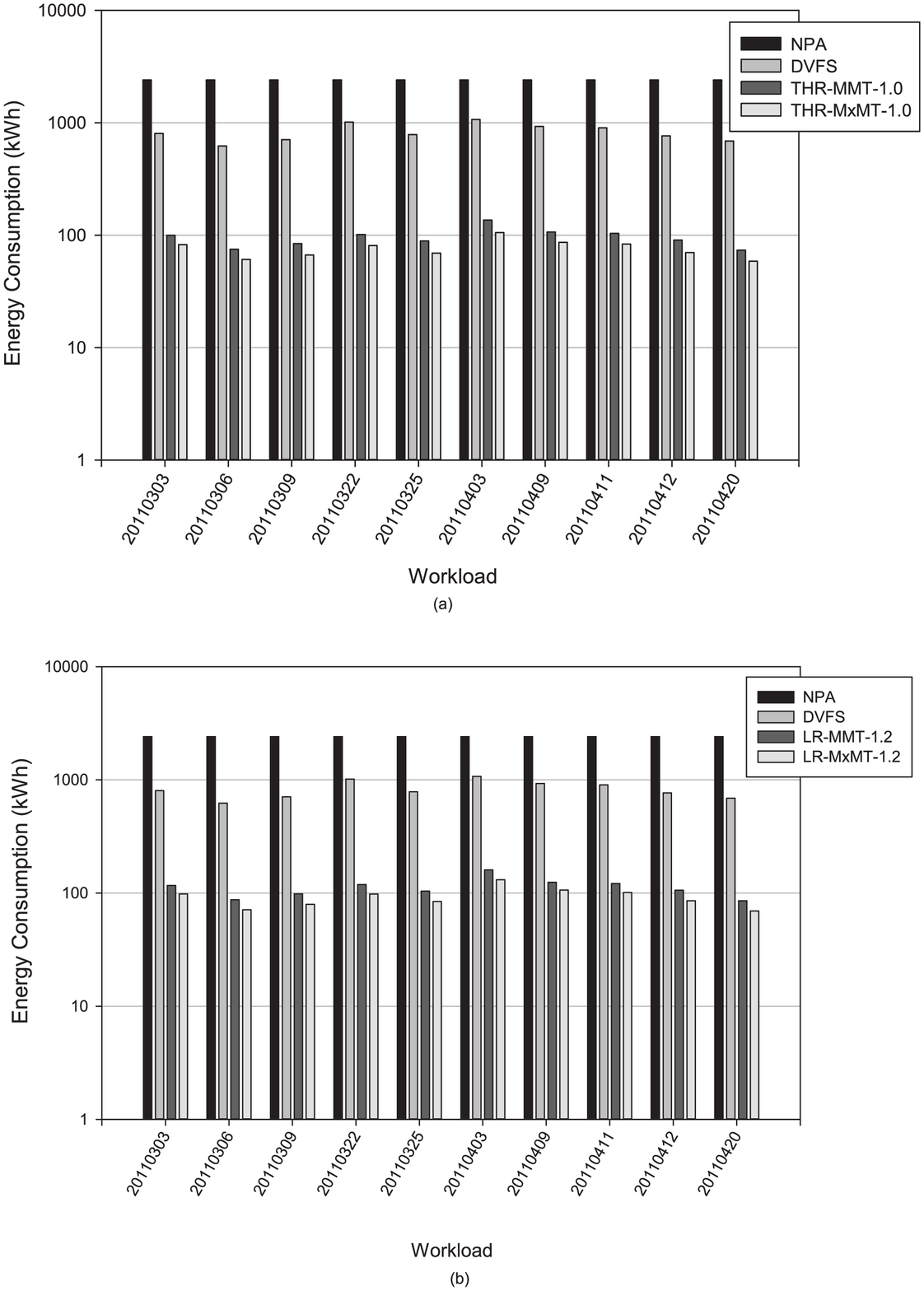}
		\caption{Result Comparison of Energy Consumption (a) Maximum Migration Time with THR 1.0 (b) Maximum Migration Time with LR.}
		\label{F2EnergyTHR1LR}       
	\end{center}
\end{figure*}

\clearpage

With the combination of our proposed algorithm the policies denoted as: THR-MxMT-1.0, THR-MxMT-0.8, IQRMxMT-1.5, MAD-MxMT-2.5, LRR-MxMT-1.2 and LR-MxMT-1.2. Simulation results had been considered  for all six algorithm combinations. Our proposed algorithm outperforms in terms of results compared with to the prior work. We measured the Energy consumption, SLATAH, and VM Migration. \\

\subsection{Energy Consumption}
Fig. 5 (a) and (b) shows the energy consumption of our proposed MxMT algorithm along with the combination of IQR and LRR. The proposed IQR-MxMT-1.2 reduced the energy consumption by 19\% on average when compared to IQR-MMT-1.2. On the other hand LRR-MxMT-1.2 saves over 18\% on average of energy consumption compared to LRR-MMT-1.2. \\

The proposed MAD-MxMT-2.5 and THR-MxMT-0.8 have reduced energy consumption by over 18\% and 19\% respectively on average. The simulation result of MAD-MMT-2.5 and MAD-MxMT are shown in Fig. 6 (a) and another simulation results of THR-MMT-0.8 and THR-MxMT-0.8 are shown in Fig. 6 (b). \\

Given the simulation result shown in Fig. 7 (a), our proposed MxMT algorithm reduced over 20\% of energy consumption on average compared to the previously proposed algorithm. Next, the proposed LR-MxMT-1.2 policy cut-off around 18\% energy consumption on average compared to LR-MMT-1.2 policy as shown in Fig. 7 (b). \\

\subsection{SLA Time Per Active Host}

Our experimental result shows that the SLA time per active host has greatly increased. As Fig. 8 (a) IQR-MxMT-1.5 shows an average of 43\% SLATAH, while the average  was 5\% in IQR-MMT-1.5. Similar results had been observed for the LRRMMT-1.2 and LRR-MxMT-1.2 policies as shown in Fig. 8 (b). For MADMMT-2.5 and MAD-MxMT-2.5 policies SLATAH found average 5\% and 46\% respectively as illustrated in Fig. 8 (c). On the other hand, for THR-MMT-0.8 and THR-MxMT-0.8 policies, an average 5\% and 40\% SLATAH had been observed respectively which are shown in Fig. 8 (d), and it points the SLATAH increment compared to the prior work. \\

On average, 27\% and 83\% SLATAH were found for THR-MMT-1.0 and THR-MxMT-1.0 policies which are illustrated in Fig. 9 (a). Fig. 9 (b) depicted SLATAH for LR-MMT-1.2 and LR-MxMT-1.2 policies. LR-MMT-1.2 produced an average of 4\% SLATAH and LR-MxMT-1.2 produced an average of 41\% SLATAH. \\

\subsection{Virtual Machine Migration}

Our experimental result shows a great reduction of VM migration which is significant for our proposed algorithm. Compared to the prior work, VM migration had decreased by 95\%, 94\%, 95\% and 96\% for IQR-MxMT-1.5, LRR-MxMT-1.2, MAD-MxMT-2.5 and THR-MxMT-0.8 respectively. The simulation result is shown in Fig. 10 (a), (b), (c) and (d). \\

Similarly, the number of VM migration had decreased for THR-MxMT-1.0 and LR-MxMT-1.2 by 94\% and 93\% as illustrated in Fig. 11 (a) and (b). The number of VM migration was reduced because we chose a heavy VM to migrate first, if the host had been considered to be overloaded.

\section{Conclusion}
\label{cons}
From the simulation results presented in this paper, we summarize that our proposed VM selection algorithm is more energy efficient compared to the prior work. On average, our proposed VM selection algorithm saves 19\% of energy cost. On the other hand, the proposed algorithm reduces over 94\% VM migration while SLATAH had increased. This is due to the migration of VM which had occupied a maximum memory. In this paper we have presented a simulation which is based on the evaluation of VM selection algorithm for dynamic VM consolidation. We have described our proposed method for VM selection. Next, we have formulated our method and had implemented our method as an algorithm. Furthermore, we have tested our proposed algorithm in a simulated cloud data center environment under a real life workload. Finally, we have presented an analysis of our simulation result which had been obtained from the experiments.

\section*{Acknowledgment}

This work is supported by the Malaysian Ministry of Education \ under \ the \ Fundamental Research Grant Scheme  FRGS/02/01/12/1143/FR.

\clearpage
\begin{sidewaysfigure*}
	\begin{center}
		\includegraphics[width=0.8\paperwidth]{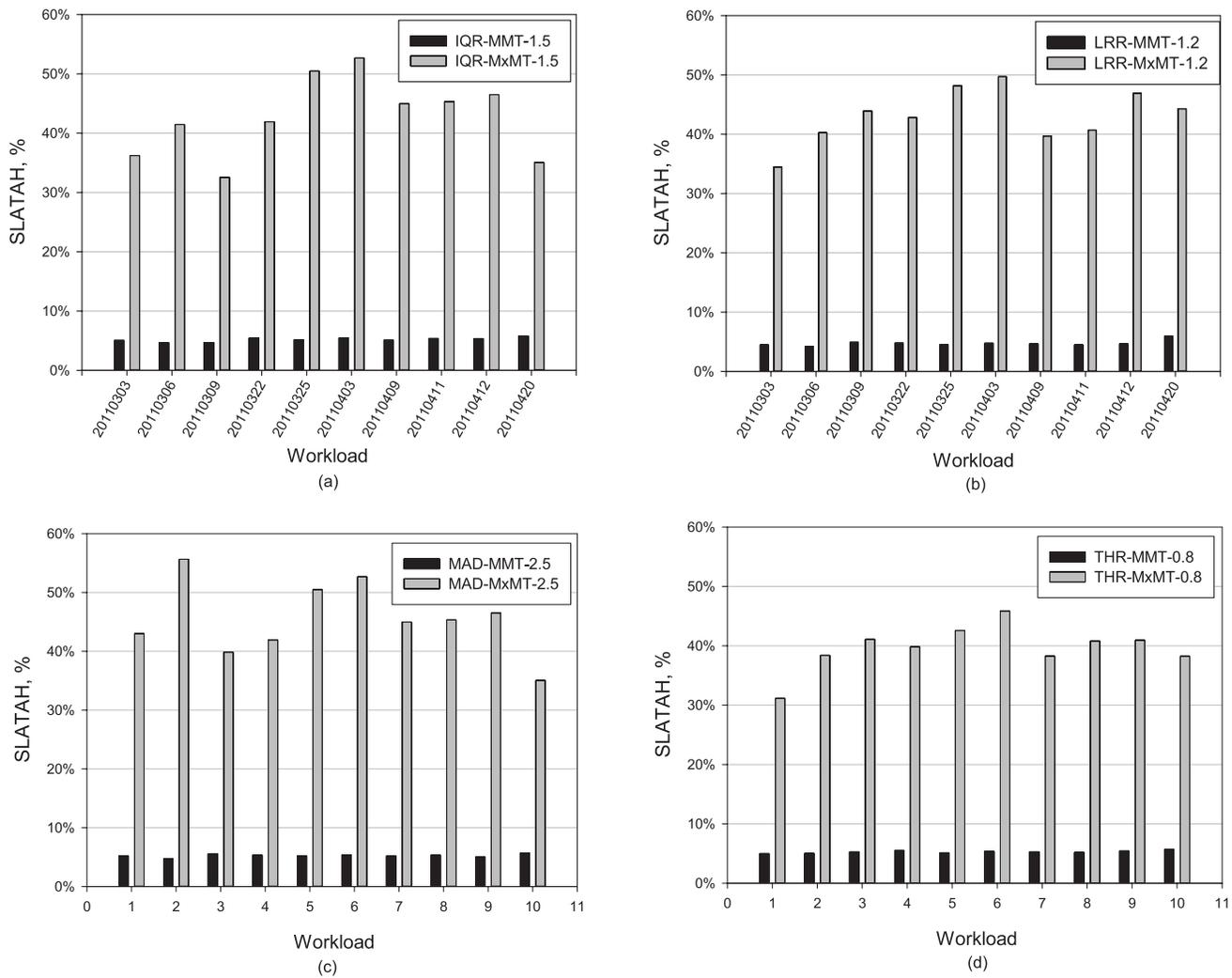}
		\caption{Result Comparison of SLA Time per Active Host (a) Maximum Migration Time with IQR (b) Maximum Migration Time with LRR (c)Maximum Migration Time with MAD (d) Maximum Migration Time with THR 0.8.}
		\label{F2SLATAHIQRLRRMADTHR}       
	\end{center}
\end{sidewaysfigure*}

\clearpage

\begin{figure*}
	\begin{center}
		\includegraphics[width=0.8\textwidth]{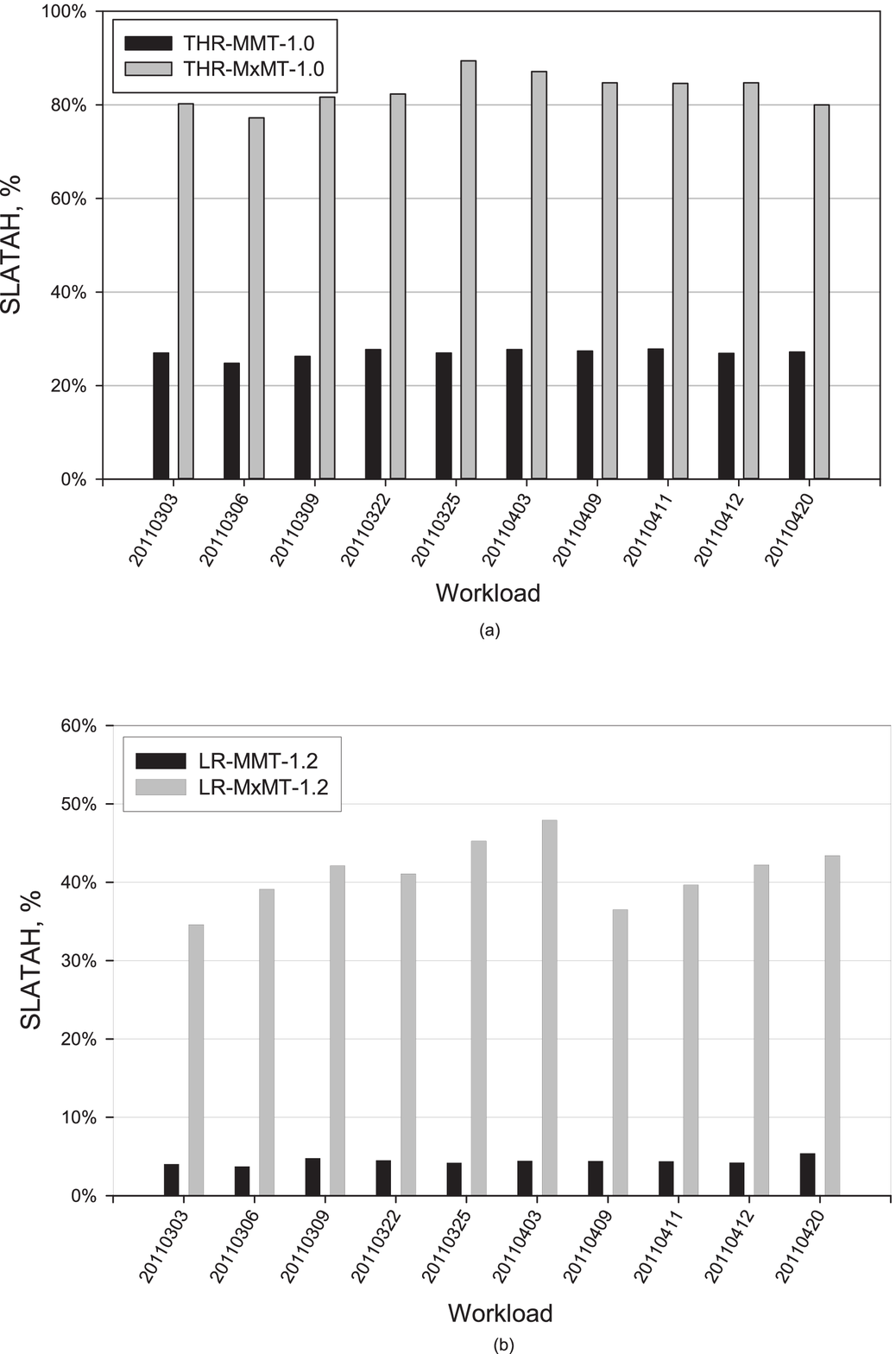}
		\caption{Result Comparison of SLA Time per Active Host (a) Maximum Migration Time with THR 1.0 (b) Maximum Migration Time with LR.}
		\label{F2SLATAHTHR1LR}       
	\end{center}
\end{figure*}

\clearpage

\begin{sidewaysfigure*}[ht]
	\begin{center}
		\includegraphics[width=0.8\textwidth]{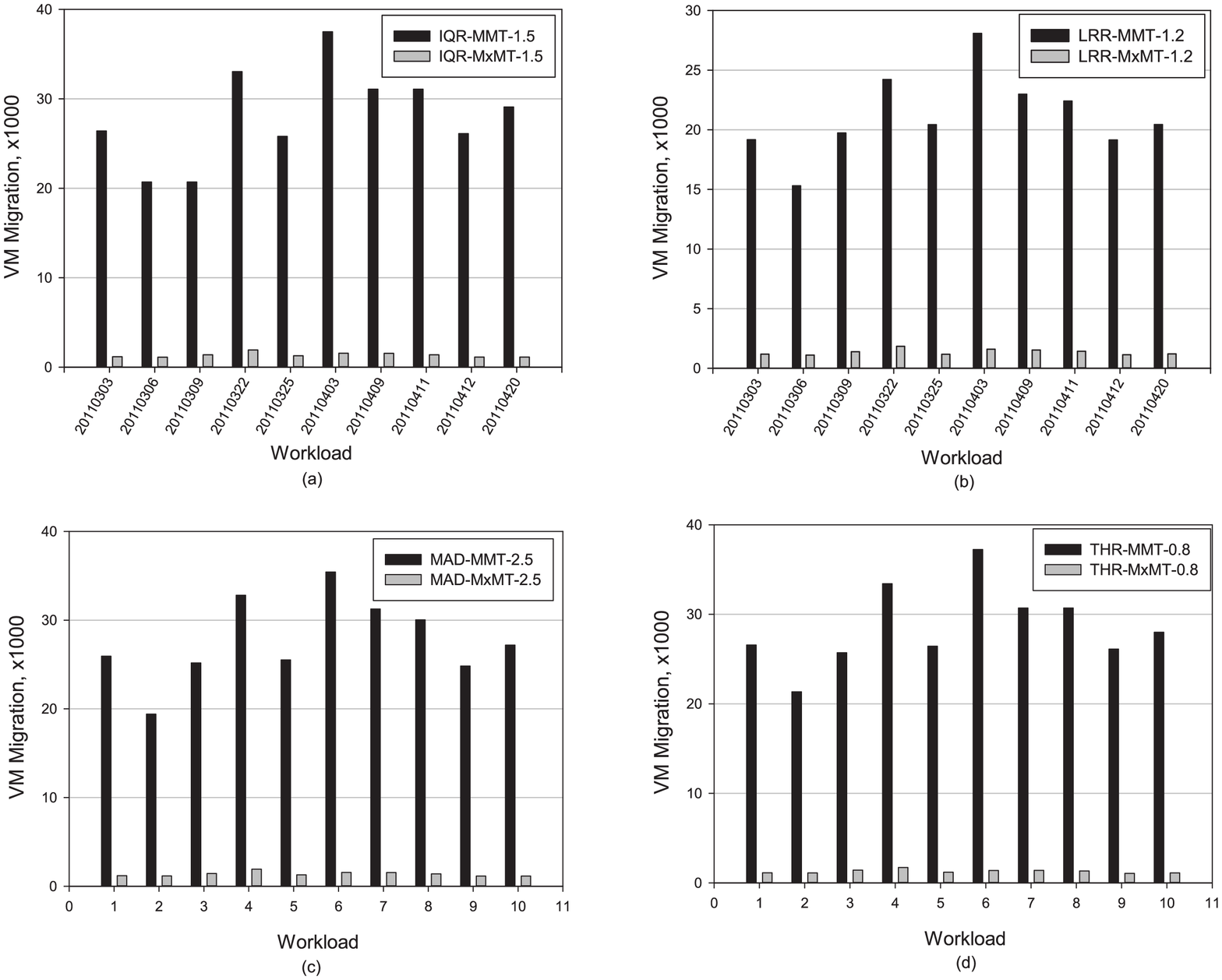}
		\caption{Result Comparison of VM Migration (a) Maximum Migration Time with IQR (b) Maximum Migration Time with LRR (c)Maximum Migration Time with MAD (d) Maximum Migration Time with THR 0.8.}
		\label{F2VMMigIQRLRRMADTHR}       
	\end{center}
\end{sidewaysfigure*}

\begin{figure*}
	\begin{center}
		\includegraphics[width=0.8\textwidth]{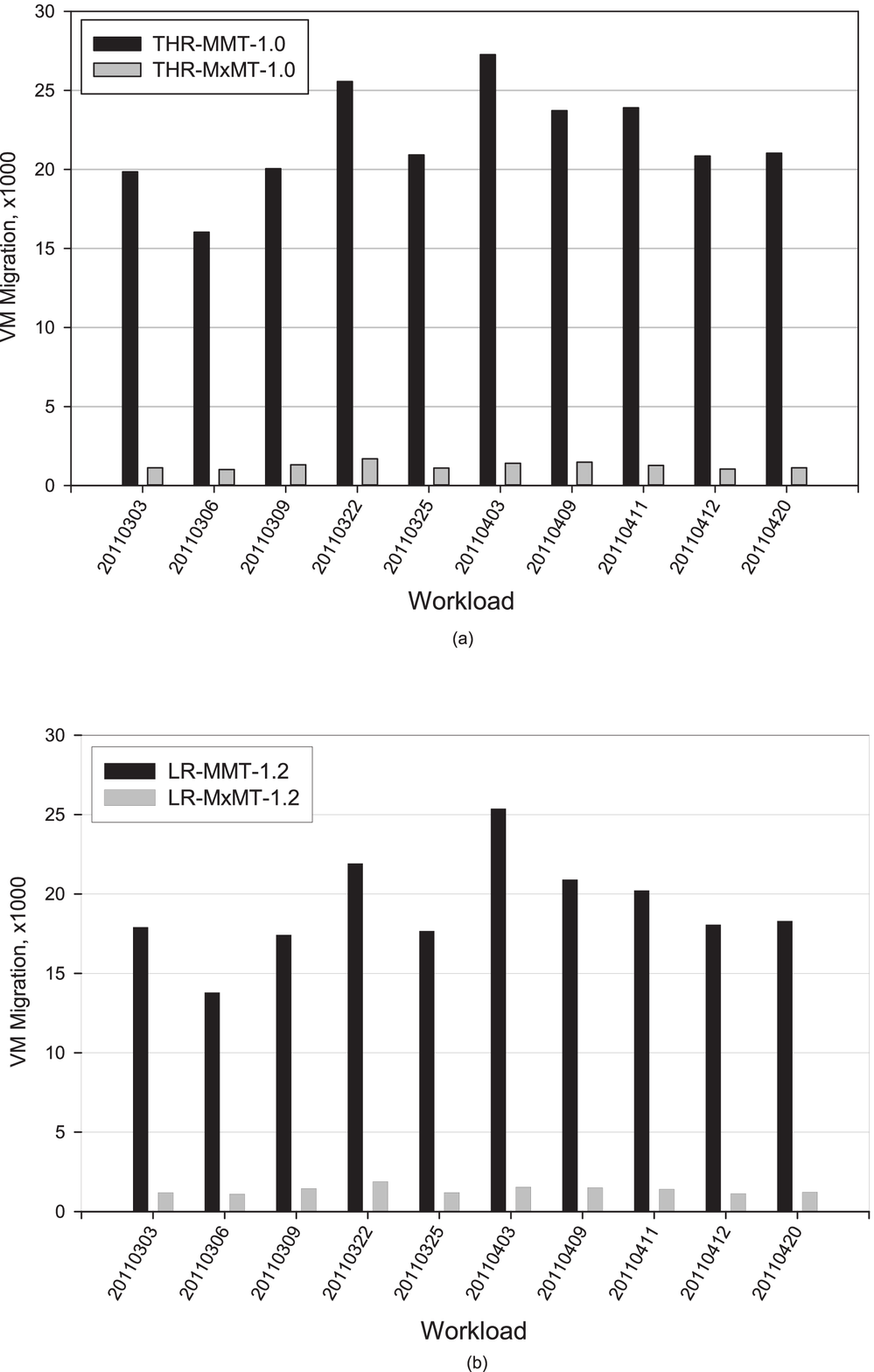}
		\caption{Result Comparison of VM Migration (a) Maximum Migration Time with THR 1.0 (b) Maximum Migration Time with LR.}
		\label{F2VMMigTHR1LR}       
	\end{center}
\end{figure*}

\clearpage

\ifCLASSOPTIONcaptionsoff
  \newpage
\fi

\end{document}